\documentclass[12pt]{article}
\usepackage{epsfig}
\usepackage{graphicx}

\newcommand{\be}{\begin{equation}}

\newcommand{\ee}{\end{equation}}

\newcommand{\eq}[1]{(\ref{#1})}
\newcommand{\fig}[1]{Fig.~\ref{#1}}

\newcommand{\cit}[1]{[\ref{#1}]}

\begin{document}

\title{\bf Strain induced stabilization of stepped Si and Ge surfaces near (001)}

\author{ V.~B.~Shenoy, C.~V.~Ciobanu and L.~B.~Freund \\
Division of Engineering, Brown University, Providence, RI
02912}

\date{\small \today}

\maketitle

\begin{abstract}

We report on calculations of the formation energies of several [100] and [110]
oriented step structures on biaxially stressed Si and Ge (001) surfaces. It is
shown that a novel rebonded [100] oriented single-height step is strongly
stabilized by compressive strain compared to most well-known step structures. We
propose that the side walls of ``hut''-shaped quantum dots observed in recent
experiments on SiGe/Si films are made up of these steps. Our calculations
provide an explanation for the nucleationless growth of shallow mounds, with
steps along the [100] and [110] directions in low- and high-misfit films,
respectively, and for the stability of the (105) facets under compressive strain.
\end{abstract}

\newpage
Strain induced self-assembly is  actively being pursued as a technique for the
fabrication of nanoscale electronic devices and memories that have the potential
to bring higher speed to information processing and higher areal and volumetric
capacity to information storage. In the past few years, significant advances
have been made on both the technological and fundamental aspects of self
assembly. On the technological side, it has become possible to prepare regular
spatial arrays of nanostructures.

 Several basic issues concerning the
physical mechanisms involved in the different stages of the formation and
morphological evolution of the nanostructures are only now becoming clear. The initial stages of
epitaxial nanostructure formation remains least well understood. Of particular interest here, recent experimental work [\ref{sutter2000}-\ref{rastelli2001}] has
revealed that (105) oriented quantum dots in SiGe/Si films grow from shallow
precursor mounds whose sidewalls are made up of widely spaced steps that are
oriented in the [100] or [110] directions for low or high Ge compositions,
respectively.  These observations cannot be understood on the basis of a
competition between surface energy increase involved in creating the walls of the
dots and the strain energy reduction through elastic relaxation. Such a competition
would lead to a nucleation barrier for the formation of the islands, while the
experiments clearly show that the stepped mounds emerge as a natural instability
without any such barrier [\ref{sutter2000},\ref{tromp2000}].

Recently we have demonstrated that nucleationless growth of stepped mounds can
be understood by including the physics of surface steps, in particular the
dependence of their formation energy on strain and their interactions
\cit{shenoy2002}. In this letter, we calculate the parameters that characterize step
 formation and interactions in order to understand the stabilization of [100] or
 [110] steps on (2$\times$1) reconstructed Si and Ge (001) surfaces at different levels of
 biaxial strain. The strain dependence of the step formation energy is
determined by the disposition of the atomic bonds in the vicinity of the
step-edge; if there is significant rebonding such that these bonds are stretched
(compressed) relative to the bonds in the bulk material, a compressive (tensile)
mismatch  stress tends to lower their formation energy \cit{xie1994}. The
interactions between steps include the repulsive dipolar interactions
\cit{marchenko1980}, which depend on the local step density, and the non-local
attractive interactions due to the force monopoles induced by the mismatch
strain \cit{tersoff1995}. The surface energy of a stepped surface that is
oriented at a small angle $\theta$ with respect to the (001) direction can  be
written by including the step formation and dipolar interaction energies as
\be
\gamma(\theta,\epsilon) =  \gamma_0(\epsilon)\cos\theta + (\beta_1 + \tilde{\beta}_1\epsilon)|\sin\theta| + \beta_3\frac{|\sin\theta|^3}{\cos^2\theta},
\label{se}
\ee
 where $\epsilon$ is the biaxial surface strain, $\gamma_0$ is the surface
 energy of the (001) surface, $\beta_1$ is the formation energy of  a step,
 $\tilde{\beta}_1$ is a measure of the sensitivity of the formation energy of a
 step to strain and $\beta_3$ characterizes the strength of the step-step dipolar
 interactions. Since the attractive monopole induced interactions favor a
 stepped surface over a flat surface, stepped mounds can grow without any
 nucleation barrier if the surface energy of their sidewalls becomes lower than
 that of the (001) surface \cit{shenoy2002}. It can be seen from \eq{se} that this indeed happens
 for stepped surfaces with orientations in the range $|\theta| < \theta_c$,
 where $\theta_c \approx \sqrt{-\beta(\epsilon)/\beta_3}$, if the condition
 $\beta(\epsilon) \equiv \beta_1 + \tilde{\beta}_1\epsilon < 0$ is satisfied.
  In the present work we show by means of atomistic simulations
 that this condition is satisfied for the mismatch strains of interest.

  In the [110] direction, we will only consider the double-height
 B-type (DB) steps \cit{chadi1987}; these steps are known to be lower in energy
 than the combination of alternating single-height  SA and SB  steps for vicinal
 angles that are more than about 1$^0$ - 2$^0$. In the [100] direction we consider
 two kinds of steps: (1) zig-zag SA+SB$^{[100]}$ steps proposed by Chadi
 \cit{chadi1987} and experimentally observed by Wu {\em et.al.} \cit{wu1993} and
 (2) straight steps on which the dimer rows of the adjoining terraces arrive at the
 steps at
 an angle of 45$^0$. If we focus attention on the family of (10n) surfaces to
 which the (105) surface belongs, there are two distinct low-energy step
 structures that can be obtained by removing every other atom from the step-edge
 in order to reduce the number of dangling bonds, as shown in \fig{steps}. The key
 difference between the two structures is that in one of them, which we call
 single-height rebonded (SR), there are atoms at the edge of the step
 bonded to atoms on the adjoining terraces, while this type of rebonding is absent
 in the single-unrebonded (SU) step (refer to \fig{steps}).
The structure of the (105)
 facet made up of SU and SR steps correspond to the reconstructions for this
 surface proposed by Mo {\em et. al.} \cit{mo1990} and Kohr and Das Sarma
 \cit{khor1997}, respectively.

We determine the step formation and interaction energies using the following
procedure:  the surface energies of various stepped surfaces are first
determined as a function of applied biaxial strain for both the (10n) and (11n)
families as shown in \fig{gamma}.  The step formation and interaction energies
are then extracted from the surface energy $\gamma(\epsilon,\theta)$ using
\eq{se}.  The strain dependent step formation energies for various step
structures calculated using the Tersoff (T3) \cit{tersoff1988} potential are shown in \fig{gamma} for Si and
Ge \cit{monopole}. In all cases, the step formation energies can be reasonably
approximated by a linear relation assumed in \eq{se} with small deviations that
can be attributed to non-linear effects, particularly at larger strains.
Calculation of step energies using the Stillinger-Weber (S-W) \cit{stillinger1985} potential also showed almost linear
dependence of the formation energies with strain. The parameters $\beta_1$ and
$\tilde{\beta}_1$, obtained by using a linear fit to $\beta(\epsilon)$,
along with the dipolar interaction strength $\beta_3$ are given in Table I.

If we focus attention on the [100] oriented steps in Si, we find that the
formation energies of the SR steps are lower than those for the SU and
SA+SB$^{[100]}$ steps for compressive mismatch strains. This can be understood
by analyzing the bond lengths  at the step-edge (refer to \fig{steps}); since
there are several stretched bonds in the SR structure, we expect its energy to
be lower in compressive strains than the SU structure. Although the zig-zag
SA+SB$^{[100]}$ steps are energetically lower than the SR steps in the absence
of strain, \fig{gamma} shows that a modest amount of compressive strain (0.3 \% in Si)
stabilizes the SR steps over these steps. Table I also shows that while the
formation energies of the SR steps on an unstrained surface are larger that the
corresponding values for the DB steps, the strain sensitivity $\tilde{\beta}_1$
for the SR steps  is about 50\% larger in both Si and Ge. This implies that the
SR steps are preferred over the DB steps at large values of compressive strain,
with one key difference between Si and Ge. Using both the S-W and T3 potentials,
we find that the formation energies of the SR steps fall below the energies for DB
steps when strains become more than about 1\% and about 4\% in Si and Ge,
respectively (refer to the insets in \fig{gamma} and Table I).

The above observations have important implications for the growth of stepped
islands observed in recent experiments [\ref{sutter2000}-\ref{rastelli2001}]. If
we assume that the results for Si are indicative of the trends in low-misfit
films ($\epsilon$ $\approx$ 1\%-2\%), it is clear that the [100] SR steps should
be observed during early stages of island growth since their formation energies
are lowest among all the cases that we have considered. While we are not aware
of a direct experimental observation confirming this prediction, our picture is
consistent with the experiments of Sutter and Lagally \cit{sutter2000} who find
that in SiGe films with 25\% Ge content, the stepped mounds are made up of [100]
oriented monolayer steps. Experiments that can identify the surface structure of
these stepped mounds will be invaluable in resolving the issue. In distinct
contrast to the low-misfit films, experiments on pure Ge \cit{vailionis2000} and
Si-capped Ge \cit{rastelli2001} islands show that the stepped mounds are made up
of [110] oriented steps \cit{scount}. These observations can also be understood
on the basis of the formation energy of the SR and DB steps in Ge shown in
\fig{gamma}. Here, the SR steps do not become favorable until mismatch strain
becomes close to 4\% and 5\% in the T3 and S-W potentials respectively; since
the strains in these mounds should be less than the mismatch between Si and Ge
(4.2\%), the observation of [110] oriented steps are in agreement with our
calculations.

While a negative step creation energy would lead to formation of surfaces with
closely spaced steps, repulsive interactions increase as  the spacing between
the steps become smaller. This competition leads to an optimum slope, which
can be determined using \eq{se} as $\theta^* \approx
\sqrt{-\beta(\epsilon)/3\beta_3}$ \cit{shenoy2002}. The existence of such a slope is indeed
confirmed by the surface energies for Ge shown in \fig{gamma}, where at 4\%
strain the (105) and (115) surfaces are seen to be the optimum facets in the
(10n) and (11n) directions respectively. In the case of the (11n) surfaces, the
(115) surface corresponds to closest spacing of DB steps (the (114) and (113)
surfaces are reconstructed so that the steps lose their identity).  In the
presence of compressive strains, the (105) surface with  SR steps has a
smaller surface energy than both the high index (10n) surfaces shown in
\fig{gamma} and the low index (103) surface.  It can also be seen from \fig{gamma}
that the surface energy of the (105) surface in Ge lies below the energy  of
the (115) surface throughout the range of interest, while it falls below the
energy of the (117) at a strain of about 3\%. In the case of Si, we find a
similar trend with the (105) surface falling below the (117) surface close to
1\% . These observations indicate that the (105) facet made up of SR steps is
stabilized by compressive strains in both low- and high-misfit SiGe films. On
the other hand, the (105) surface with the SU steps proposed by Mo {\em et.al.} is
stabilized by tensile  rather than compressive stresses since the bonds at the
step-edge are in compression (refer to \fig{steps}). While the results in the
present work are obtained using empirical potentials , we have verified by means
of {\em ab initio} simulations that the atoms at the edges of the SR steps in
the (105) surface are indeed in a stretched state. The details of these calculations
will be published elsewhere.

In summary, the nucleationless growth of stepped mounds can be understood on the
basis of strain dependence of [100] and [110] oriented steps on (001) surfaces.
The rebonding at the edge plays an important role in stabilizing stepped
surfaces in the presence of strain. The competition between repulsive step
interactions and strain induced lowering of step formation energies leads to
optimum low-energy  orientations such as (105). Further work that analyzes the effect of
surface segregation on step formation energies will be invaluable in gaining deeper insight into early stages of quantum dot formation in SiGe and other alloys.

\vspace{1cm}
\noindent{\em Acknowledgments :}{\small \setlength{\baselineskip}{11pt}{ The research support of the National
Science Foundation through grant CMS-0093714 and the Brown University MRSEC
Program, under award DMR-0079964, is gratefully acknowledged. }}

\vspace{1cm}
\newpage
\noindent{\large{\bf{References}}}

\begin{enumerate}

\item \label{sutter2000}
P.~Sutter and M.~G.~Lagally,  {\it Phys. Rev. Lett.} {\bf 84}, 4637 (2000).

\item \label{tromp2000}
R.~M.~Tromp, F.~M.~Ross and M.~C.~Reuter,  {\it Phys. Rev. Lett.} {\bf 84}, 4641 (2000).

\item \label{vailionis2000}
A.~Vailionis, B.~Cho, G.~Glass, P.~Desjardins, D.~G.~Cahill and J.~E.~Greene, {\it Phys. Rev. Lett.} {\bf 85}, 3672 (2000).

\item \label{rastelli2001}
A.~Rastelli, M.~Kummer and H.~von Kanel,  {\em Phys. Rev. Lett.} {\bf 87}, 6101 (2001).

\item \label{shenoy2002}
V.~B.~Shenoy and L.~B.~Freund,  A continuum description of the energetics and evolution of stepped surfaces in strained nanostructures, {\em J.~Mech.~Phys.~Sols.} (in press); A copy of this article can be found at {\em http://arXiv.org/abs/cond-mat/0203514}.

\item \label{xie1994}
Y.~H.~Xie, G.~H.~Gilmer, C.~Roland, P.~J.~Silverman, S.~K.~Buratto, J.~Y.~Cheng, E.~A.~Fitzgerald, A.~R.~Kortan, S.~Schuppler, M.~A.~Marcus and P.~H.~Citrin, {\bf 73} 3006 (1994); C.~Roland, {\it MRS Bulletin} {\bf 21}, 27 (1996).

\item \label{marchenko1980}
V.~I.~Marchenko and A.~Ya.~Parshin,  {\it Sov. Phys. JETP} {\bf 52}, 129 (1980).

\item \label{tersoff1995}
J.~Tersoff, Y.~H.~Phang, Z.~Y.~Zhang and M.~G.~Lagally,  {\it Phys. Rev. Lett.} {\bf 75}, 2730 (1995).

\item \label{chadi1987}
D.~J.~Chadi, {\it Phys. Rev. Lett.} {\bf 59},  1691 (1987).

\item \label{wu1993}
F.~Wu, S.~G.~Jaloviar, D.~E.~Savage and M.~G.~Lagally, {\it Phys. Rev. Lett.} {\bf 71},  4190 (1993).

\item \label{mo1990}
Y.~W.~Mo, D.~E.~Savage, B.~S.~Swartzentruber and M.~G.~Lagally, {\it Phys. Rev. Lett.} {\bf 65}, 1020 (1990).

\item \label{khor1997}
K.~E.~Khor and S.~Das Sarma, {\em J.~Vac.~Sci.~Technol.~B}, {\bf 15} 1051 (1997).

\item \label{tersoff1988}
J.~Tersoff, {\em Phys. Rev. B}, {\bf 38}, 9902 (1988); {\em Phys. Rev. B}, {\bf 39} 5566 (1989).

\item \label{stillinger1985}
F.~H.~Stillinger and T.~A.~Weber, {\em Phys. Rev. B}, {\bf 31}, 5262 (1985); K.~Ding and H.~C.~Andersen, {\em Phys. Rev. B}, {\bf 34} 6987 (1986).

\item \label{monopole}
In the case of SR steps there should be an additional term in \eq{se} that accounts for interaction of the monopoles
 due to the discontinuity of surface stress at the step-edge. We have ignored this term in our analysis of vicinal surfaces with $\theta>$ 3$^0$ since the magnitude of this term becomes significant for small vicinal angles ($\theta<$ 1$^0$).

\item \label{scount}
From the STM images of the stepped mounds in \cit{rastelli2001} we have inferred that the [110] oriented steps are of the double-height type.

\end{enumerate}

\newpage
\begin{table}[h]
{Table I: Step formation energies $\beta_1$, their derivatives with respect to
strain $\tilde{\beta}_1$, and
the step-step dipolar interaction coefficient $\beta_3$
for Si and Ge (with values for Ge enclosed in parentheses) calculated using the Stillinger-Weber (S-W) and Tersoff (T3) potentials. In order to allow a direct comparison of the formation energies of different step structures, $\beta_1$ is measured relative to the formation energy of the DB step calculated using the T3 potential. All quantities are expressed in meV/ \AA$^2$.}
\begin{center}
\begin{tabular}{|l| l r|| l r|| c|| c | }
\hline \hline
 &\ \ \ \ \ \ \ \ \ \ \ SR& &\ \ \ \ \ \ \ \ \ \ \ DB & & SU & SA+SB$^{[100]}$  \\
 & S-W & T3 &  S-W & T3 & T3 & T3  \\
\hline \hline

$\beta_1$ &19.9(29.9)&5.8(10.1)&12.1(14.8)&0.0(0.0)& 46.3(38.6)& 5.98(0.96) \\
$\tilde{\beta}_1$  &741(738)  &1046(781)&440(459)  &667(501)&   -93(-89)& 523(385)\\
$\beta_3$ &169(186)  & 217(143)&239(189)  &307(229)&      2(11)& 585(382) \\
\hline \hline
\end{tabular}
\end{center}
\label{fulltable}
\end{table}
\newpage
\vspace{-6cm}
\begin{figure}
\center{\bf{Figure Captions}}
\vspace{1cm}
\caption{The structure of the [100] oriented single-height rebonded (SR) step (top) and the unrebonded (SU) step (bottom) on a (2$\times$1) reconstructed (001) surface. The dashed line denotes the edge of the step, where the dimer orientation undergoes a 90$^0$ rotation. Atoms are colored according to the number of dangling bonds(b) {\em before} surface and step-edge reconstructions: red=2b, green=1b, blue=0b.
The numbers represent the stretching (in \% ) of some of the bonds relative to the bond-length in the bulk for the Si(109) surface relaxed using the T3 potential. Note that the rebonding of the atoms across the step edge leads to stretching of several bonds in the
SR step.  }
\label{steps}
\end{figure}
\begin{figure}
\caption{Surface energy of stepped
surfaces (in meV/ \AA$^2$) consisting of [100]SR(red) and
[110]DB(black) steps as a function of the biaxial strain
$\epsilon$ for Ge, computed with the T3 potential.  The vicinal
angles and surface orientations are indicated in the figure.
The insets show the strain dependent formation 
energy $\beta(\epsilon)$ (in meV/\AA$^2$)
for the three types of steps, DB(black squares), SR(red circles) and 
SA+SB$^{[100]}$ (blue triangles) in Si and Ge.}
\label{gamma} 
\end{figure}
\newpage
\begin{figure}
\begin{center}
\bf {Figure 1: Shenoy, Ciobanu and Freund}
\vspace{2cm}
\includegraphics{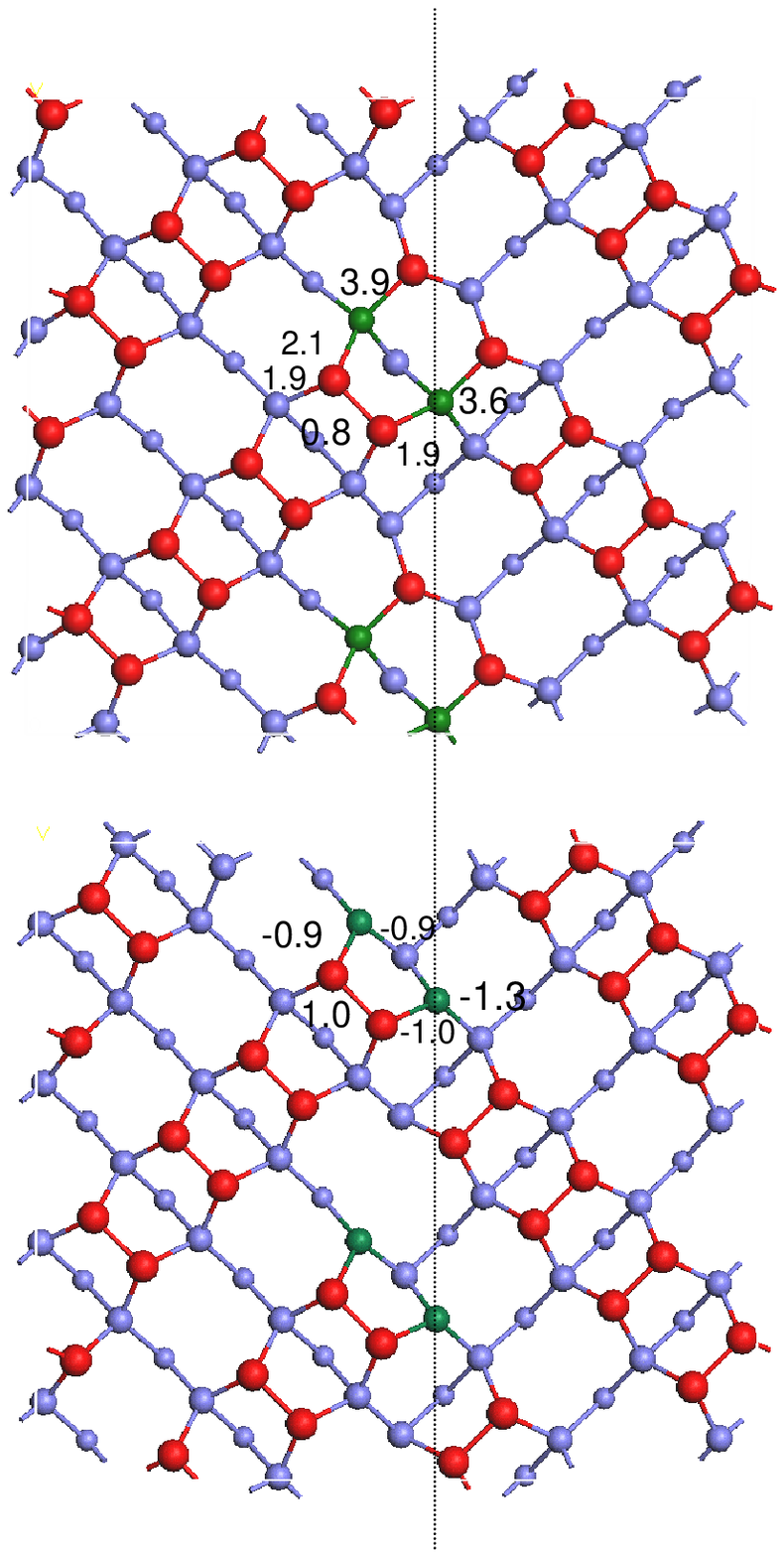}
\end{center}
\end{figure}
\newpage
\begin{figure}
\begin{center}
\bf {Figure 2: Shenoy, Ciobanu and Freund}
\vspace{2cm}
\includegraphics{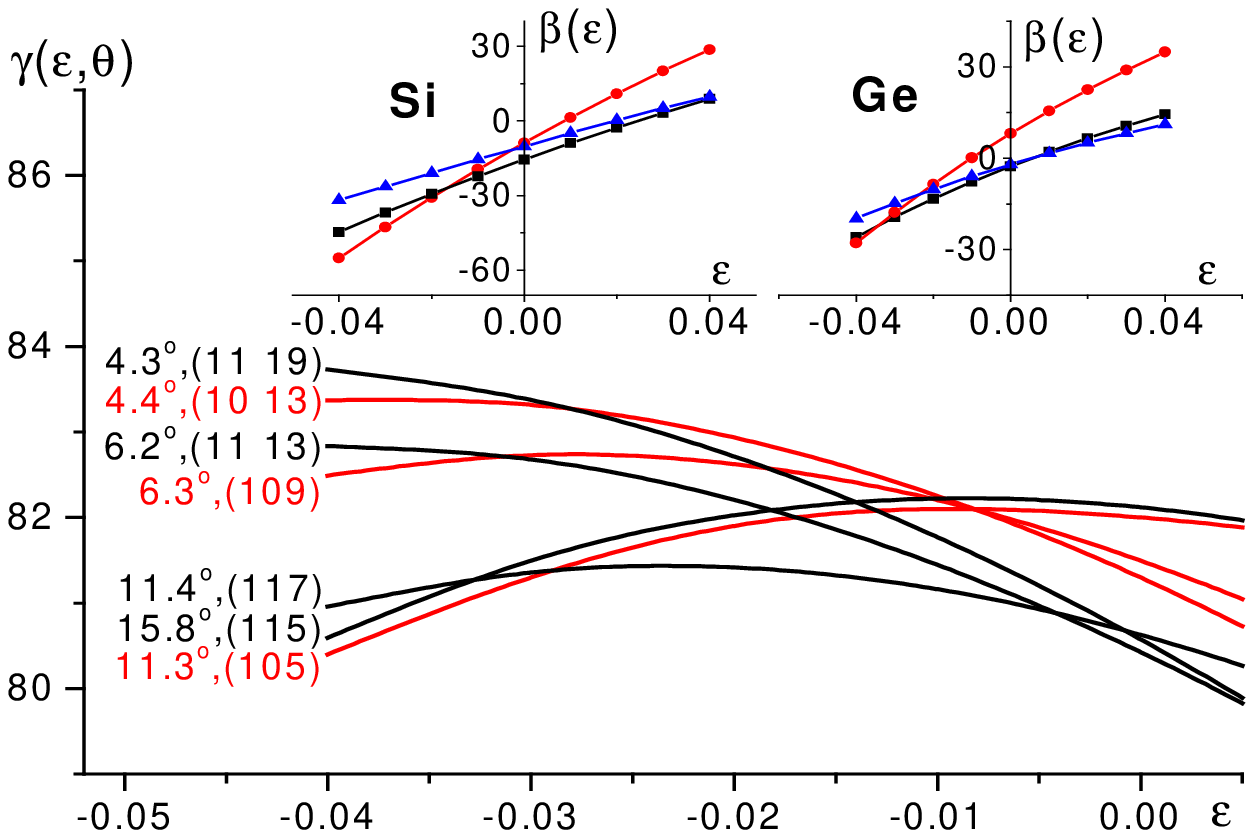}
\end{center}
\end{figure}

\end{document}